# WEB SERVICES DISCOVERY AND RECOMMENDATION BASED ON INFORMATION EXTRACTION AND SYMBOLIC REPUTATION

Mustapha AZNAG[1], Mohamed QUAFAFOU[1], Nicolas DURAND[1] and Zahi JARIR[2]

[1]Laboratory of Information and Systems Sciences, Aix-Marseille University, France
`{mustapha.aznag,mohamed.quafafou,nicolas.durand}@univ-amu.fr`
[2]LISI Laboratory, FSSM, University of Cadi Ayyad, Morocco
`zahijarir@ucam.ac.ma`

## ABSTRACT

*This paper shows that the problem of web services representation is crucial and analyzes the various factors that influence on it. It presents the traditional representation of web services considering traditional textual descriptions based on the information contained in WSDL files. Unfortunately, textual web services descriptions are dirty and need significant cleaning to keep only useful information. To deal with this problem, we introduce rules based text tagging method, which allows filtering web service description to keep only significant information. A new representation based on such filtered data is then introduced. Many web services have empty descriptions. Also, we consider web services representations based on the WSDL file structure (types, attributes, etc.). Alternatively, we introduce a new representation called symbolic reputation, which is computed from relationships between web services. The impact of the use of these representations on web service discovery and recommendation is studied and discussed in the experimentation using real world web services.*

## 1. INTRODUCTION

Web services[1] [1] are defined as a software systems designed to support interoperable machine-to-machine interaction over a network. They are loosely coupled reusable software components that encapsulate discrete functionality and are distributed and programmatically accessible over the Internet. They are self contain, modular business applications that have open, internet-oriented, standards based interfaces [7]. Different tasks like matching, ranking, discovery and composition have been intensively studied to improve the general web services management process. Thus, web services community has proposed different approaches and methods to deal with these tasks. Empirical evaluations are generally proposed considering different simulation scenarios. Nowadays, we are moving from web of data to web of services as the number of UDDI Business Registries (URBs) is increasing. Moreover, the number of host that offer available web services, through specific engines like Axis[2], is significantly increasing. Consequently, the WSDL files describing services can be crawled and parsed automatically. Such files contain different

---

[1] http://www.w3.org/standards/webofservices/
[2] http://axis.apache.org/axis/

kind of information like textual descriptions, simple and/or complex types, attributes, etc. (see Section 3 for more details).Contrary to simulation based approaches, real world services inherit the complexity of the real world. With the increasing number of published Web services providing similar functionalities, it's very tedious for a service consumer to make decision to select the appropriate one according to her/his needs. Therefore mechanisms and techniques are required to help consumers to make the best choice. In addition, web services use natural language descriptions (ambiguity of sentences), they are multi-languages and cross-domains. In this context, the representation of web service becomes a major challenge. Moreover, the quality of representations has an important impact on services registries that provide mechanisms for web services discovery, composition and recommendation. These tasks have been intensively studied in the literature [4, 16, 17, 24, 9, 2, 12]

This paper studies different representations of web services and compares them considering the following important tasks: discovery and recommendation. The first representation is centred on textual descriptions of services and their available functions. It is produced from the web service descriptions and enriched by integrating the descriptions of operations offered by services and the structural elements of WSDL, for instance simple and/or complex types, attributes. The previous representation is traditional and well known by the community of Web services. We consider this representation as a **baseline representation(B)**for a web service. We introduce two advanced representations that refine B considering semantic aspect of text and we show their interest inthe experiments (see Section 7):

- **Rules based text tagging (RBTT)**: a lot of services have very detailed descriptions, especially when they offer several operations with their own descriptions. The main question is how to recognize significant parts or entities in the text description and how to use the filtered information to describe the web service?

All the previous representations are produced from the web service description, i.e, WSDL content, which represents the provider point of view. Our goal is to infer a web service representation from the description of its context (neighbour services).

- **Symbolic reputation (SR):** web services reputation is generally a numeric quantity computed from user feedback [17]. The representation we consider here is qualitative as it is produced from the symbolic descriptions of other web services. Consequently, such a description describes the relationships of a service with others and does not describe the service itself.

We introduce in this paper methods that compute the different representations to evaluate their interest for discovery and recommendation of web services. We also describe a general architecture of a System developed and used in the experimentation.
This paper is organized as follows. Section 2 provides an overview of related work. Section 3 details the WSDL parsing and the information extraction. Section 4 presents the different representations of web services. The new proposed representations web services (RBTT and SR) are discussed respectively in Section 4.2 and 4.3. Section 5 presents the general architecture of our system. Section 6 is dedicated to the web service discovery and recommendation system. Finally the experimental results are discussed in Section 7 before concluding.

## 2. RELATED WORK

In this section, we briefly discuss some of the research works related to discovering Web services. Although various approaches can be used to locate and discover Web services on the web, we have focused our research on the service reputation and discovery problems. Every web service associates with a WSDL document that contains the description of the service. A lot of research efforts have been devoted in utilizing WSDL documents [16, 28, 29, 25, 12, 26].Dong et al. [16] proposed the Web services search engine Woogle that is capable of providing Web services similarity search. However, their engine does not adequately consider data types, which usually reveal important information about the functionalities of Web services [27]. Liu and Wong [25] apply text mining techniques to extract features such as service content, context, host name, and name, from Web service description files in order to cluster Web services. They proposed an integrated feature mining and clustering approach for Web services as a predecessor to discovery, hoping to help in building a search engine to crawl and cluster non-semantic Web services. Elgazzar et al. [12] proposed a similar approach, which clusters WSDL documents to improve the non-semantic web service discovery. They take the elements in WSDL documents as their feature, and cluster web services into functionality based clusters. The clustering results can be used to improve the quality of web service search results.

In [30], the authors proposed an architecture for Web services filtering and clustering. The service filtering mechanism is based on user and application profiles that are described using OWL-S (Web Ontology Language for Services). The effectiveness of the filters is based on a clustering analysis that compares services related clusters. The objectives of this matchmaking process are to save execution time and to improve the refinement of the stored data. Another similar approach [31] concentrates on Web service discovery with OWL-S. The OWLS is first combined with WSDL to represent service semantics before using a clustering algorithm to group the collections of heterogeneous services together. Finally, a user query is matched against the clusters, in order to return the suitable services. Nevertheless, the creation and maintenance of ontologies may be difficult and involve a huge amount of human effort [28, 29].

Web service reputation represents a mechanism relying on feedbacks provided by consumers/software agent to measure Web service trustworthiness. It is modelled as a vector of aggregate consumers/software agent's ratings for a web service. Also various rating feedbacks are aggregated to derive a service provider's reputation. Generally feedback is composed from collected quality of service (QoS) information acquired from execution monitoring and those that require consumer's intervention like price or accuracy that cannot be monitored. According to the QoS information published and a consumer preference including required QoS metrics, the QoS registry will calculate an overall rating for each web service that matches the consumer's search request. Then the consumer will select the web service with the highest rating [17]. Most of proposed reputation approaches use a central registry to collect and share consumer's feedbacks. Since this central architecture is subject of failure, other works based on peer-to-peer web services are proposed to deal with decentralized reputation mechanism [32]. A service provider that provides satisfactory service may get incorrect or false ratings from unfair or malicious raters. One of challenging problem is protecting web service reputation from these incorrect inputs. Some mechanisms have been proposed to detected and deal with dishonest feedbacks by using dedicated monitoring agents to filter consumers evaluation [33], or collaborative filtering techniques based on peer to peer solution [34]. In the literature, a rating of a service is a vector of attribute values. The computed reputation rating may be a binary value (trusted or untrusted), a scaled integer (e.g. 1-10), or on a continuous scale (e.g., [0, 1]). Therefore the satisfaction level of web services is generally a normalized numerical value, representing quantitative reputation, used for dynamic services ranking and selection.

Maximilien and Singh [3, 35] proposed a multi-agent based architecture where agents assist in quality-based service selection using an agency to disseminate reputation and endorsement information. Each proxy agent is autonomous but also collaborates with other agents to collect other opinions and therefore maximizes its information to improve its decision-making. Liu, Ngu and Zeng [36] proposed an algorithm about how to combine different QoS metrics to get a fair overall rating for a web service. The proposed reputation can be defined as the average ranking given to the service by end users. Majithia et al. [5] consider ratings of services in different contexts and a coefficient (weight) is attached to each particular context. This coefficient reflects its importance to a particular set of users. Based on this coefficient, they propose a method to compute the reputation score as weighted sum of ratings for a service. Wish art et al. [37] introduce an aging factor for the reputation score, which is applied to each of the ratings for a service. Based on these two works, Xu et al. [17] propose to calculate reputation score as the weighted average of all ratings of a service received from consumers, including an inclusion factor representing the weight attached to each of the ratings for the service. Trust and reputation mechanisms are closely related. Web service reputation can be considered as an aggregation of evaluation for a service from consumers/software agent, while web service trust represents a personalized and subjective opinion reflecting a web service [38]. Currently, several researches in the area of trust and reputation topic are considered. We cite for instance the work [33] that proposes an algorithm to analyse the trustworthiness for each consumer, which ultimately facilitates the web services selection process considering feedbacks reported by trusted users than others.

Web service community has been introduced to easy-of-use web service discovery. It can be viewed as a mediator that holds the meta-data and registry information about its member services and represents domain-specific knowledge [39]. Among Community-based approach, we present for instance the work [34] that has proposed a community-based service selection approach based on super agents. These agents share their information about services they have interacted with, which is useful for other agents to make effective selection of services. This in order to maintain communities and build community-based reputation for a service based on the opinions from all community members that have similar interests and judgement criteria.

These efforts have an implicit hypothesis, which considers that web service reputation is a numerical quantity. As far as we know, it's the first time we break the quantitative reputation hypothesis. This paper avoids this hypothesis considering the qualitative aspect of the reputation and describes how to compute and use it.

## 3. WSDL DOCUMENT PARSING AND INFORMATION EXTRACTION

The Web Service Description Language (WSDL) is an XML-based language, designed according to standards specified by the W3C that provides a model for describing web services. It describes one or more services as collections of network endpoints, or ports. It provides the specifications necessary to use the web service by describing the communication protocol, the message format required to communicate with the service, the operations that the client can invoke and the service location. Two versions of WSDL recommendation exist: the 1.1[3]version, which is used in almost all existing systems, and the 2.0[4] version which is intended to replace 1.1. These two versions are functionally quite similar but have substantial differences in XML structure.

To manage efficiently web services descriptions, we extract all features that describe a web service from the WSDL document and store them into a relational database. We recognize both WSDL versions (1.1 and 2.0). During this process, we proceed in three steps (see Figure 1). The first step consists of checking availability of web service and validating the content of WSDL document. The second step is to get the WSDL document and read it directly from the WSDLURI to extract all information of the document. In this step we describe the features to extract from the WSDL document: (1) the name, the documentation and the version of the WSDL, (2) WSDL types used by messages to transmit information between web services. Data types are often specified using a XML Schema Definition (XSD). We extract all kind of elements and types that can be as simple or complex types as a set of elements and/or attributes, and (3)a set of services declared in the WSDL document. For each service we extract the name, the documentation and a set of endpoints. Then, for each endpoint we extract the name, the address (which defines the connection point to web service. It is typically represented by a simple HTTP URL) and the binding(Name, Type, Style, Transport protocol). The binding specifies the interface as well as defining the SOAP binding style (RPC/Document) and transport SOAP protocol. The interface defines the operations to be performed for a web service, and the messages that are used to perform the operation. We also extract for each operation the name, the documentation, the input and the output parameters. The input/output parameters can be referred to the previously extracted types/elements. Finally, the third step is dedicated to save the extracted WSDL features into a database. The extracted information will be used during the generation of representations(**B, RBTT** and **SR**). Before presenting the methods for calculating representations, we discuss some text-processing standard used thereafter.

---

[3]http://www.w3.org/TR/wsdl

[4]http://www.w3.org/TR/wsdl20/

1. **Tag removal**: This step removes all HTML tags, CSS components, symbols(punctuation, etc.).
2. **Splitting and remove a stop words**: Some terms are composed by several words separated by a capital letter; we use therefore regular expression to extract these words. To illustrate, the application of the regular expression ([A-Z][a-z]+) on this string "GetAll Country Currencies Response" produces' Get', 'All', 'Country', 'Currencies' and 'Response'. Furthermore, to extract the potential content words, we remove all the stop words. Finally, the potential content words for the previous example are 'Country' and 'Currencies'.
3. **Word Stemming**: In this step we use the Porter Stemmer [21] to remove words, which have the same stem. Words with the same stem will usually have the same meaning. For example, 'computer', 'computing' and 'compute' have the stem 'comput'.

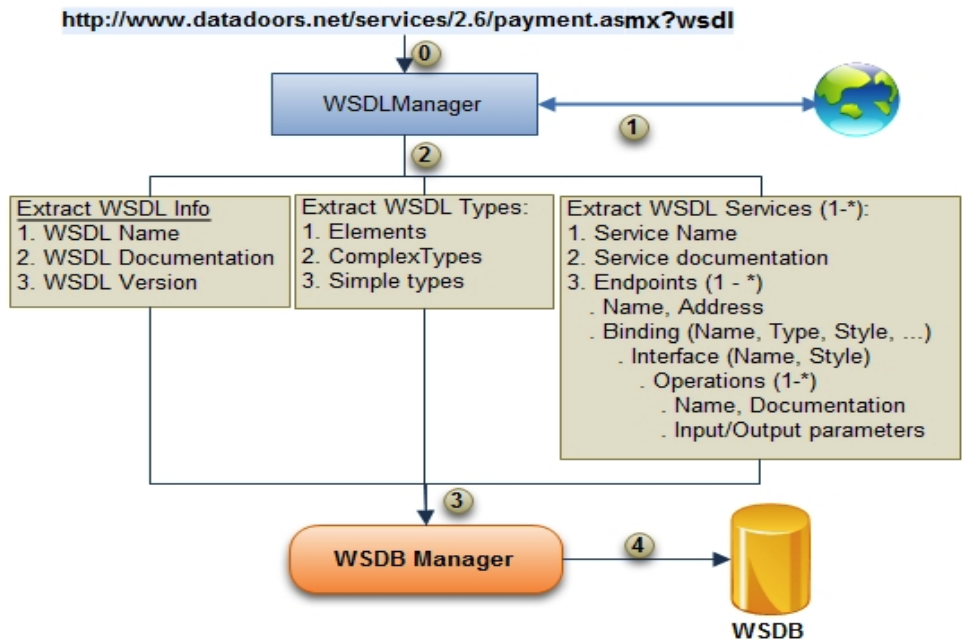

Figure 1. WSDL document parsing and information extraction

## 4. REPRESENTATIONS OF WEB SERVICES

In this section, we present a generation method of the traditional representation of web service and we introduce two new representations. Let us note that the generated representations are vectorial.

### 4.1. Baseline representation (B)

A web service can be described by a textual description extracted from WSDL document or given by its provider when publishing in the UDDI. The current UDDI registry only allows searching web services by their textual description. The first representation is centred on textual descriptions of services and their offered functions. This is produced from the web service descriptions and enriched by integrating the descriptions of operations offered by services. Let us remark that the major disadvantage is that most web services have a poor or an empty description. To complete this representation, we added all information described by WSDL types. The types are used by messages to transmit information between web services. Consequently, WSDL types are good features to describe the functionality of a service and are the most informative element in WSDL document. For this reason, we extract all type names (elements, complex types, simple types, attributes, enumerations) and we apply the textual processing (see Section 3 -steps 2 and 3) to produce a set of words. Thus, we use the obtained set of words to construct the new representation and we consider it as a baseline representation (B) for a web service.

### 4.2. Rules Based Text Tagging of web services descriptions (RBTT)

A lot of services have very detailed descriptions, especially when they offer several operations with their own descriptions. The main question is how to recognize significant parts or entities in the text description and how to use the filtered information to describe the web service? Our approach consists in the definition of extraction rules to identify, extract and annotate relevant multi-word terms from web service descriptions. This approach has been already used for biological data [19]. The processing steps(Tokenization, Part-Of-Speech tagging, Extraction and output generation) of Information Extraction have been implemented as modules using the *Lingua Stream* platform[5]. *Lingua Stream* [20] is an integrated experimentation environment targeted to researchers in natural language processing (NLP). Let us note that we used *Tree Tagger*[6] for the *Part-Of-Speech* tagging step. The extracted information is given in a form of a XML file. In the context of the web service descriptions, we have defined a set of rules into *Prolog* reflecting the *Definite Clause Grammar* (*DCG*) to recognize 3types of information: web service names (*namews*), purposes of the web services(*purpose*), and the domain of utilization (*domain*). Let us remark that we do not use patterns in the sense of Information Extraction that is without a prior on the form of the expressions.

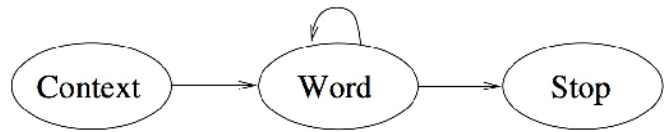

Figure 2. Structure of rules

Figure 2 presents the structure of the rules. From a *'context'*, an expression(generally a multi-word term, a nominal phrase) is recognized until a stop phrase is encountered. The context is a set of *'trigger'* words. The stop phrases can be words, symbols, verbs, punctuation, etc. They depend on the rule type.

---

[5] http://www.linguastream.org/

[6] http://www.ims.uni-stuttgart.de/projekte/corplex/TreeTagger/DecisionTreeTagger.html

Currently, the identification of the context has been done manually, however an automatic learning of the context can also be considered. We have manually detected special phrases for each type of information on an excerpt of the corpus. We have noted the corresponding trigger words and the stop phrases. The common stop phrases are the trigger words, some punctuation symbols and the relative pronouns. The current rule base consists of 35 rules – 13 for namews,16 for purpose, and 6 for domain. Let us note that different cases can be expressed by the same rule. Let us take some examples.

**Example 1:** The web service description is "*CAPTCHA-image web service for serving and validating CAPTCHA-images* ". We obtain this result:

*<ws><id>WSID 172</id>*
*<namews>****CAPTCHA-image****</namews></ws>*

**Example 2:** The web service description is "*The ICD9 coding system is an international classification system which groups related disease entities and procedures for the purpose of reporting statistical information. The system is widely used to for medical billing* ". We obtain this result:

*<ws><id>WSID 29</id>*
*<domain>****medical billing****</domain></ws>*

In the example 1, the term "CAPTCHA-image" is extracted by using the following set of rules:
*Webservname (type:namews..name:N) --> start, nws(N).*
*start --> @lemma:'.' ; ls_lookupToken(_,_,startPara).*
*nws(N) -->ls_token(N,tag:X,_), end, {not(member(X,[dt]))}.*
*nws(N) -->ls_token(N1,_), nws(N2), {concat(N1,N2,N)}.*
*end --> @lemma:web, @lemma:service ; @lemma:webservice.*

 The trigger phrase is the start of a paragraph ("*startPara* ") or the start of a new sentence. The rules "*nws* " allow the system to recognize the multiword terms. The end phrase is here *Web Service* or *Web Service*  (with or with out capitals). Let us remark that the rule "*nws* " avoid the recognition of a single determiner ("*dt* ") just before the end phrase.

Our system differs from the classical approaches. From Information Extraction method point of view, simple declarative extraction rules are designed, making the implementation process "light and quick". The rules are domain-specific, but by no means corpus-specific. Moreover, our method is endogenous: no resources such as knowledge base or dictionary are needed at the beginning. There sources are constructed on the fly – the system learns new terms (which can be new terms in the domain or missing) to be used later or in other corpora and/or in other text mining applications.

All the previous representations are produced from the web service description, i.e, WSDL content, which represents the provider point of view. Our goal is to infer a web service representation from the description of its context (neighbour services).

## 4.3. Symbolic reputation model (SR)

The reputation of a web service reflects a common perception of other web services or customers towards that service. In other words, it aggregates the web service ratings given by consumers.

Typically, a reputation would be built from a history of ratings. Several reputation systems have been proposed in the literature [17, 3, 5, 13]. However, feedbacks from users are not easily available and Hidden Markov Models (HMM) were used to predict the reputation of a service provider [10]. In [8], the authors propose an algorithm for generating locally calculated reputation ratings from a semantic network. This effort has an implicit hypothesis, which considers that web service reputation is a numerical quantity. As far as we know, it is the first time we go beyond the quantitative reputation by proposing a symbolic reputation notion. In fact, our work considers the qualitative aspect of the reputation and describes how to compute and use it. In this paper, we consider that the numeric reputation of a service $s$, denoted $R_{num}(s)$, results from customer's feedback and computed as the weighted average of all ratings the service received from customers [17]:

$$R_{num}(s) = \frac{\sum_{i=1}^{N_s} s^i * I^{d_i}}{\sum_{i=1}^{N_s} I^{d_i}} \qquad (1)$$

Where $N_s$ is the number of ratings for the service $s$, $s^i$ is the $ith$ service rating, $I \in [0,1]$ is the inclusion factor tacking into account the dimension time (smaller $I$ means that the more recent ratings have a larger impact in the reputation score), and $d_i$ is the age of the $ith$ service rating.

Before computing symbolic reputation, we explain the structure of the web service space by modelling relationships between services. Web services dependency graph was used to represent the structure of the web services space, i.e. web services relationships, by generating a network depending on the Input / Output of each web service. This induced network is then used for web service composition [23,2]. The general approach consists of searching first for web services operations dependencies using the I/O parameters of their operations then the web services dependency graph is induced. For seek of simplicity, we consider here that a dependency occurs between two services when they offer at least two dependent operations: a service requires (resp. provides) data from (resp. to) another service. of course, other dependency definitions can be used without affecting our general symbolic reputation computing approach.

### Definition 4.3.1 Operations dependency:

Let $F = \{f_1, f_2, ..., f_n\}$ be the set of operations offered by all web services published in the UDDI. $InF(f_i)$, $OutF(f_i)$ represent respectively the inputs and the outputs of the operation $f_i$, $a \in [0,1]$ a given threshold and $sim$ the similarity function.

The operation $f_j$ depends on the operation $f_i$, denoted $f_i \to f_j$, if and only if $\forall p \in InF(f_j), \exists q \in OutF(f_i) : sim(p,q) \geq a$. $sim(a,b) = 1 - NGD(a,b)$ is the featureless similarity factor computed between words $a$ and $b$ using Normalized Google Distance (NGD) [22] as a featureless distance measure between words.

### Definition 4.3.2 Web services dependency graph:

Let $S = \{s_1, s_2, ..., s_n\}$ be the web services space containing all services published in the UDDI, we define the dependency graph as the directed graph $G = (S, V)$ where $V = \{(s_i, s_j) \in S \times S, \exists f_i \in s_i, \exists f_j \in s_j : f_i \to f_j\}$. We note $f \in s$ when the service $s$ offers the operation $f$.

This construction of the direct web services dependency graph is needed to compute the symbolic reputation. Our model of symbolic reputation is based on random walks models. A random walk is a sequence of nodes in a graph constructed by the following process: select a random starting node in the graph, then move to select a neighbour of this node at random and so forth. The random walk analysis has been applied to different fields as a fundamental model for random processes in time [11]. More particularly, a formal model was proposed in [6] to allow the computation of the reputation of a web page using the web hyperlink structure. We have adapted this model to compute the symbolic reputation of web services.

We consider the constructed dependency graph. The probability of visiting a service $s$ for term $t$ at step $n$ of the walk, denoted $P^n(s,t)$, is defined as follows:

$$P^n(s,t) = (1-d) \sum_{q \to s} \frac{P^{n-1}(q,t)}{O(q)} + \begin{cases} \dfrac{d}{N_t} & if \quad t \in VR(q) \\ 0 & Otherwise \end{cases} \qquad (2)$$

The previous probability is the core of the algorithm that computes the symbolic reputation. In fact, suppose that with the probability $d$ the random walker jumps into a service uniformly chosen at random from the set of services that contains the term $t$. In this context, the probability that a random walker visits a service $s$ in a random jump is $\dfrac{d}{N_t}$ ( $N_t$ denotes the total number of services on the UDDI that contains the term $t$ ) if the vectorial representation $VR(q)$ of service $q$ contains the term $t$ and it is zero otherwise. The probability that the walker visits the service $s$ at step $n$ after visiting one among its parent service $q$ is $\dfrac{1-d}{O(q)} P^{n-1}(q,t)$ where $P^{n-1}(q,t)$ denotes the probability that the walker visits the service $q$ for term $t$ at step $n-1$ and $O(q)$ denotes the number of services that require data from $q$ (parents of $s$ in the dependency graph). Let us denote $VR(s)$, the vectorial representation of service $s$. Algorithm 1 gives the details on the computation of the symbolic reputation of a given service.

## 5. GENERAL ARCHITECTURE AND UDDI EXTENSION

The aim of our implementation is to extend a UDDI by introducing the different approaches previously discussed. Our system supports service consumers to discover web services that meet their requirements. It also recommends other services to customers using symbolic reputation. In addition, our platform allows enriching the UDDI register automatically by collecting Web services published in others UDDI registries. In our implementation, services or agents (dynamic Web services) of our system reside in a remote server that consumers know a priori. Our system enables service providers to publish their services with QoS (price, availability, response time, . . .) and consumers to find web services that meet their functional requirements and/or QoS requirements. Figure 3 gives an overview of the general architecture by outlining the different components and their relationships. The general architecture of our system consists of the following components:

---

*ALGORITHM 1: SYMBOLIC REPUTATION COMPUTING*

---

REQUIRE:

- d: The random surfer jump.
- k: Maximum number of iteration.
- s: A given service.

ENSURE: $SR(s)$: Symbolic reputation of the service s.

1:    $SR(s) = \varnothing$

2:    $VR(s)$ = get Vectorial Representation(s)

3:    FORALL term $t \in VR(s)$ DO

4:    $P(s,t) = d / N_t$

5:    ENDFOR

6:    FOR $l = 1, 2, .., k$ DO

7:    IF $l < k$ THEN

8:    $d' = d$

9:    ELSE

10:    $d' = 1$

11:    ENDIF

12:    FOR every path $q_l \to ... \to q_1 \to s$ of length $l$ and every term $t$ in $VR(q_l)$ DO

13:    $P(s,t) = 0$ if term $t$ has not been seen before

14:    $P(s,t) = P(s,t) + [(1-d)^l / \prod_{i=1}^{l} O(q_i)(d' / N_t)]$

15:    ENDFOR

16:    ENDFOR

17:    FOR every term $t$ with $P(s,t) > 1 / N_t$ DO

18:    $SR(s) = SR(s) \cup t$

19:    ENDFOR

20:    RETURN $SR(s)$

---

- **UDDI Registry**: It allows service providers to publish their web services.
- **UDDI Manager**: It contains two interfaces **Service Publisher** and **Service Finder**. **Service Publisher** offers all the necessary operations to service providers to manage their services in the UDDI registry. This is the only possible way to interact with the UDDI registry. **Service Finder** allows finding services according to functional requirements. It can also provide information that describes a web service published in the UDDI registry.
- **Discovery Agent**: It receives consumer's requests, finds the web services that meet their requirements and sends the response to the customer. It also collects feedback rating from customers expressed as a score reflecting their level of satisfaction after interacting with the discovered service.
- **Reputation Manager**: It allows calculating the symbolic reputation of web services.
- **Clustering Manager**: It allows creating the clusters of web services.
- **WSDL Manager**: It allows parsing, extracting features from WSDL documents and save them in the **WSDB** database.
- **WVMC rawler**: It is a dynamic web service that runs automatically to collect Web services published in other UDDI registries.
- **WSDB Manager**: It allows managing the WSDB database (CRUD - Create/Read/Update/Delete operations).

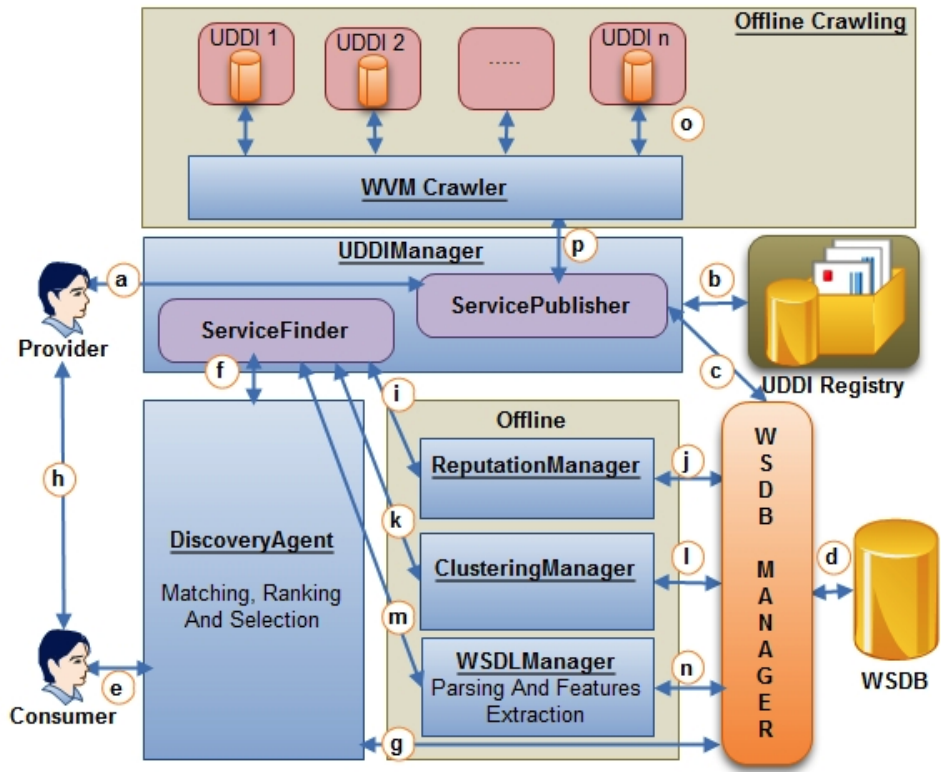

Figure 3. General architecture and UDDI extension

(a). Services register.

(b). Register / unregister services and business on UDDI Registry.

(c). Save the QoS advertisements (price, availability, response time, . . .) if are specified for services to register on database WSDB.

(d). CRUD (Create/Read/Update/Delete) operations on database WSDB.

(e). Request/Response.

(f). Find services from UDDI by functional requirements.

(g). (1) Get Qos and reputation advertisements for services that functional requirements match with user's query. (2) Send user's feedback score for web service.

(h). Bind a web service.

(i). Get services from UDDI and calculate the Quantitative/Qualitative reputation.

(j). Save the computed reputation for each service on WSDB.

(k). Get services from UDDI and construct the clusters of Web services.

(l). Save the constructed clusters on WSDB.

(m). Get the new WSDL URI published on UDDI Registry and parse it to extract all information of WSDL document.

(n). Save the features extracted by WSDL Parser on WSDB.

(o). Crawler WSDLs URI from others UDDI registries.

(p). Register the new crawled web services on UDDI registry.

## 6. WEB SERVICE DISCOVERY AND RECOMMENDATION SYSTEM

The proposed web service discovery and selection algorithm is based on the three following operators: Matching, Ranking and Selection (see Algorithm 2):

- **Matching**: Find services that meet the consumer's functional requirements. Then maintain services that meet only the non-functional requirements (QoS and/or reputation).

- **Ranking**: Calculate an overall score that combines the quality of service(availability, response time, . . .) and/or reputation score and use it to classify the web services,

- **Selection**: Select only the best services according to the classification results and the maximum number of services specified by consumer, and return them.

Algorithm 2 shows the details of the discovery process and selection of our system. The functional matching process is based on the consumer's functional requirements. The Qos matching process is based on the algorithm proposed by Maximilien and Singh [4] and improved by Xu et al. [17].A service consumer sends a service discovery request to the Discovery Agent (see Figure 3), which then contacts the UDDI Manager and WSDB Manager to find the services that meet the functional requirements (func Matching() - line 1).If no matched service is found, the Discovery Agent returns an empty result to the consumer. If multiple services match the functional requirements the discovery service contacts the WSDB Manager to retrieve the Qos attributes for candidates services previously selected. If multiple services match the Qos requirements (Qos Matching() - line 5) the discovery agent calculates the Qos score for each service. If the consumer's reputation requirement is specified, the discovery agent contacts the WSDB Manager to retrieve the reputation score for each candidate services. Then the ranking process (Ranking() - line 8) calculates the overall score and uses it to rank the candidates services. If no reputation requirement is specified the ranking process (Ranking - line 12) uses only the computed Qos score. Finally, the selection process (Select) selects nbMax services with the highest scores and returns them to the consumer (nb Maxdenotes the maximum number of service to be returned as specified by the consumer). If nbMax is not specified, one service is randomly selected from services whose Qos score (or overall score) is greater than the given threshold.

We also propose an algorithm for web services recommendation using the symbolic reputation (SR). Algorithm 3 shows the details of the symbolic reputation based recommendation. The general method is based on the following steps: (1) we use Algorithm 2 to discover services that match consumer's requirements.(2) We retrieve the symbolic reputation SR(s) for each discovered services. Then, we find all services q that have a vectorial representation $VR(q)$ such that $SR(s) \cap VR(q) \neq \varnothing$ and its cardinality is greater than a given threshold $\mathfrak{a}$. Then, the ranking process (Ranking - line 9) uses the computed Qos score and the reputation score to rank the candidates services previously selected. Finally, the selection process (Select - line 10) selects the services with the highest scores and returns them to the consumer. Only the best services according to the ranking result are recommended.

---

***ALGORITHM 2: WEB SERVICE DISCOVERY AND SELECTION***

---

**REQUIRE:**

- **fRq**: Functional requirements
- **qosRq**: Qos requirements
- **repRq**: Reputation requirements
- **nbMax**: Maximum number of services to be returned

**ENSURE: select**: A set of discovered services

---

```
1:   Service[] f Match= func Matching(fRq)
2:   select = Service[]
3:   qMatch = Service[]
4:   IFqosRq != null THEN
5:   qMatch= Qos Matching(f Match,qosRq)
6:   sRank = Service[]
7:   IFrepRq != null THEN
8:   sRank= Ranking(qMatch, qosRq, repRq)
9:   select  = Select(sRank,nbMax,"QosAndRep")
10:  RETURN select
11:  ELSE
12:  sRank=   Ranking (qos Match, qosRq)
13:  select=  Select (sRank, nbMax, "Qos")
14:  RETURN select
15:  ENDIF
16:  ELSE
17:  select=  Select(fMatch, nbMax, "Functional")
18:  RETURN select
19:  ENDIF
```

*ALGORITHM 3: SYMBOLIC REPUTATION BASED RECOMMENDATION*

**REQUIRE: ds**: Discovered web service. **S:** set of all services in the UDDI

**ENSURE: L**:  List of web services to recommend

1:   $L \neq \varnothing$
2:   $SR(ds)$ = get Symbolic Reputation(ds)
3:   **FORALL** $q \in S$ **DO**
4:   $VR(q)$ = get Vectorial Representation (q)
5:   **IF** $\left\| SR(ds) \cap VR(q) \right\| \geq a$ **THEN**
6:    $L = L \cup q$
7:   **ENDIF**
8:   **ENDFOR**
9:   $L$ = Ranking( $L$ )
10:  $L$ = Select( $L$ )
11:  **RETURN** $L$

# 7. EVALUATION

We have considered different web service sources like WebservicesX.net[7], xMethods.net[8]and seekda.com[9]. Our general UDDI architecture allows using different UDDI registries, which are used through specific wrappers. We have collected around 8,500 services. However, these services are not classified into categories. This is a real problem when we want to evaluate the usefulness of our representations using precision and recall measures. For this reason, we have considered a new web services source, i.e. service-finder.eu[10], which classifies its services using an ontology [18]. The number of services, which are classified into a category by service-finder is only reduced to only 1,647 web services. 465 services among them are not available: they are discarded. The collected web services are multi-languages. We have automated the recognition of the language and keep only those using English language. Finally, 993 web services are used.

---

[7] http://www.webservicex.net/ws/default.aspx

[8] http://www.xmethods.net/ve2/index.po

[9] http://www.seekda.com

Let us denote 'Collected WS' the number of collected web services for each category,' Available WS' the number of available services and 'Used WS' the number of used services for each category after the process of recognition of the language. See Table 1 for more details.

**Table 1.** Classification of the used web services according to the category

| Categories | Collected WS | Available WS | Used WS |
|---|---|---|---|
| Weather | 57 | 39 | 35 |
| Address Information | 433 | 316 | 301 |
| SMS | 45 | 36 | 20 |
| Currency Exchange | 27 | 23 | 18 |
| Stock | 163 | 153 | 142 |
| Payment | 47 | 23 | 15 |
| Converters | 23 | 15 | 9 |
| News | 37 | 29 | 25 |
| Jobs | 18 | 13 | 12 |
| Travel | 653 | 433 | 331 |
| Logistics Shipping | 14 | 12 | 6 |
| Multimedia Video | 10 | 10 | 8 |
| Identity Verification | 22 | 13 | 11 |
| Government | 11 | 7 | 7 |
| Mathematics | 36 | 26 | 24 |
| Translation | 51 | 34 | 29 |
| **Total** | **1647** | **1182** | **993** |

In the rest of this section, we discuss the results of the web services discovery considering the different presented representations. Finally, we discuss the web services recommendation results.

We use the two classical measures, Precision and Recall, to evaluate the performance of our approach. Precision and Recall are often used to evaluate information retrieval schemes [14]. Precision can be seen as a measure of exactness or fidelity, whereas Recall is a measure of completeness [14]. In our context, the Precision and Recall are defined as follows:

$$\mathrm{Pr}ecision \ = \ \frac{A}{B} \quad \textbf{(3)}$$

$$\mathrm{Re}call \ = \ \frac{A}{C} \quad \textbf{(4)}$$

Where $A$ is the number of relevant services retrieved, $B$ is the total number of irrelevant and relevant services retrieved, and $C$ is the total number of relevant services in the whole collection.

In order to evaluate the interest of the evoked representations, i.e., B, RBTT and SR , we have considered the following experimental protocol :

---

[10] http://www.service-finder.eu

- Discover the web services that belong to a given category. To achieve this goal, we generate a query $req_c$ defined by the name of category $c$, for example "Weather", "Address Information", "SMS", "Currency Exchange" and "Stock".

- Compute the precision and the recall for each representation. These measures compare the returned result with the services belonging to the category (see Table 1).
- Evaluate the global precision and recall for each representation.

The results, given in Table 2, show that the categories of web services are not represented at the same level of precision. In general, the representation RBTT is the best. Then we have B and in last SR. Note that the overall recall value is medium about all representations. In fact, the RBTT is the best representation for the category 'Weather'. For the other categories, we note that only the RBTT has obtained the medium values. Let us remark that B and SR representations give the lowest precisions for respectively "Currency" and "Jobs"(see Table 2). Finally, we have the category "SMS" for which we obtain medium results except for SR that gives the lowest precisions. In conclusion, this is difficult to find an efficient representation for all categories as the description of web services are generally heterogeneous. The lessons we have learned from our experimental results are:

1. The traditional representation B may be efficient for some categories but is not robust and may lead to very low precisions,
2. The symbolic reputation (SR) is not efficient for web services discovery, except for one category (i.e. 'Weather'), and consequently it will not be used for web services discovery,
3. The rules based text tagging (RBTT) is robust (i.e. the results do not vary according to the category) and gives correct precision results for all categories.

The previous results show that the symbolic reputation is not appropriate for web service discovery. In fact, it does not represent the description of a service itself, but it represents the description of the relation the service has with others. Thus, we have used the SR representation for web services recommendation (see Algorithm 3). The basic idea behind this task is to enrich the returned results during the web services discovery. For each discovered service, we search the set of web services to be recommended with these discovered services. The result of a query is not reduced only to the services that match the query, but it also contains recommended services for each discovered service (Target service).

The recommended services are ranked using the ranking process (Algorithm 3), which uses the computed Qos score and the reputation scores. We select only the services with the highest scores and return them to the consumer. Table3 shows the results for services belonging to the "Weather" category. We have evaluated manually these recommended services. The result is really significant as the recommended services generally perform the same task, i.e. belongs to the same category of the target service or may be composed with it.

**Table 2.** Web service discovery evaluation (Precision (P) and Recall (R) in %).

| Categories | B | | RBTT | | SR | |
|---|---|---|---|---|---|---|
| | P % | R% | P% | R% | P% | R% |
| Weather | 68,97 | 57,14 | 81,82 | 51,43 | 68,97 | 57,14 |
| SMS | 48,48 | 80,00 | 55,56 | 50,00 | 12,12 | 80,00 |

| Currency Exchange | 16,67 | 16,67 | 58,33 | 77,78 | 8,51 | 22,22 |
|---|---|---|---|---|---|---|
| Jobs | 04,35 | 08,33 | 57,14 | 33,33 | 6,67 | 41.67 |
| Average | 34,61 | 40,53 | 63,21 | 53,13 | 24,06 | 50,27 |

**Table 3.** Recommended web services for the category "Weather"

| ***Target service:*** |
|---|
| http://www.deeptraining.com/webservices/weather.asmx?wsdl |
| ***Recommended web services:*** |
| http://ws.cdyne.com/WeatherWS/Weather.asmx?wsdl<br>http://ws.soatrader.com/harbormist.com/0.1/WeatherService?WSDL<br>http://rightactionscript.com/webservices/nusoap/server.php?wsdl<br>http://www.deeptraining.com/webservices/weather.asmx?wsdl<br>http://ws.soatrader.com/weather.gov/0.1/ndfdXML?WSDL<br>http://ws.soatrader.com/cis.temple.edu/0.1/DOTSFastWeather?WSDL<br>http://ws2.serviceobjects.net/fw/fastweather.asmx?wsdl<br>http://ws.serviceobjects.com/fw/FastWeather.asmx?WSDL<br>http://ws.soatrader.com/webservicex.com/0.2/GlobalWeather?WSDL<br>http://ws.soatrader.com/ejse.com/0.1/Service?WSDL<br>http://ws.soatrader.com/deere.com/0.1/WeatherForecastServiceService?WSDL |

# 8. CONCLUSION

This paper revisits the traditional representation of web services using its textual descriptions and the elements of the WSDL structure as types, attributes, etc. We have proposed two new web services representations. The first one is based on semantic tagging of web services descriptions to keep only the most significant parts. Whereas, the second proposed representation, i.e., symbolic reputation is more contextual as it does not consider only the service itself, but its neighbours. The main conclusions of the experimentation with real-world web services are:

- Traditional representation (B) behaves correctly but it is not robust when we consider different concepts (categories).
- Rules based text tagging (RBTT) is more robust(i.e. the results do not vary according to the category) even if its precision is lower than traditional representations.
- Symbolic reputation (SR) is not appropriate for the web services discovery tasks but it is more efficient for the web services recommendation task.

In the future, we will continue our work on symbolic reputation for the recommendation task. We will increase the number of web services used for the experimentation. Finally, we will make available online the gate corresponding to the full implementation of our system.